\documentclass[12pt]{iopart}

\usepackage{graphicx}
\usepackage[usenames]{color}

\begin{document}

\title[Dynamics of gap  solitons in  a dipolar BEC on a 3D optical lattice]%
{Dynamics of gap  solitons in  a dipolar Bose-Einstein condensate on a three-dimensional optical lattice}

\author{P. Muruganandam$^{1,2}$\footnote{anand@cnld.bdu.ac.in} 
and S. K. Adhikari$^1$\footnote{Email: adhikari@ift.unesp.br; URL:
http://www.ift.unesp.br/users/adhikari/}}
\address{$^1$Instituto de F\'{\i}sica Te\'orica, UNESP - Universidade Estadual 
Paulista, 01.140-070 S\~ao Paulo, S\~ao Paulo, Brazil}
\address{$^2$School of Physics, Bharathidasan University, 
Palkalaiperur Campus, 
Tiruchirappalli 620024,   
Tamilnadu, 
India}

\begin{abstract} We suggest and study the   stable disk- and cigar-shaped  
gap solitons
 of a dipolar Bose-Einstein condensate of $^{52}$Cr atoms localized in 
the lowest band gap by three optical-lattice (OL) potentials along 
orthogonal directions. The one-dimensional version of these solitons of 
experimental interest confined by an OL along the dipole moment 
direction and harmonic traps in transverse directions is also 
considered. Important dynamics of (i) breathing oscillation of a gap 
soliton upon perturbation and (ii) dragging of a gap soliton by a moving 
lattice along axial $z$ direction demonstrates the stability of gap 
solitons. A movie clip of dragging of three-dimensional gap soliton 
is included. 

 \end{abstract}


\maketitle

Solitons in Bose-Einstein condensate 
(BEC) have drawn a great deal of attention recently. Experimentally,
bright solitons were created 
in attractive BEC of $^7$Li  \cite{4r}
and 
$^{85}$Rb \cite{5r}
atoms resulting from a cancellation of nonlinear attraction and 
linear dispersive effects. For a repulsive BEC, dark solitons have 
also been experimentally observed \cite{dark}.
In a  repulsive BEC on a periodic 
optical-lattice (OL)  
localized (dynamically) stable states 
 of gap solitons in the band gap
can be made \cite{10r,10r2}. 
An OL simulates the 
periodic electron-atom potential in a solid, and the study of gap solitons 
in OL and the related band-gap structure in the presence of nonlinear 
interaction 
is also of interest in condensed-matter physics \cite{bgap}.   
The mathematical model of gap solitons and atomic band-gap in a BEC is 
identical to that of 
light propagation in photonic crystals with cubic nonlinearity in presence
of photonic band-gap and this makes the study of gap solitons in a BEC of  general interest \cite{10r}. 
Later, the possibility of gap solitons in Fermi superfluid has been suggested
\cite{gapfermi}.
Gap solitons have been observed in photonic crystals \cite{gapf}.
A gap soliton of about 250 
$^{87}$Rb atoms in a nearly one-dimensional (1D) OL  \cite{11r,12r}
has  been created experimentally. 
In the experiment, the OL was subjected to acceleration, 
in order to push atoms into a gap soliton state \cite{11r}.
Other 
possibility for the creation of gap solitons may employ 
the 
phase-imprinting method \cite{create}.
 
 The alkali metal atoms used in BEC experiments have
negligible dipole moment. However, most bosonic atoms
and molecules have large dipole moments and a $^{52}$Cr
BEC \cite{pfau}, with a larger long-range dipolar interaction superposed
on the short-range atomic interaction, has been
realized. Thus one can   study a
dipolar BEC (DBEC)
with variable  short-range interaction \cite{pfau}
using a Feshbach resonance \cite{fesh}. 
The stability of a DBEC depends not only 
on the  atomic interaction, but also on  trap geometry \cite{pfau,jb,Dutta2007}.
In a disk configuration the  dipolar interaction is
repulsive and a DBEC is more stable; while, in a cigar 
configuration the  dipolar interaction is
attractive and a DBEC is less stable due to  
collapse instability
\cite{pfau,Yi2000}.
The controllable
short-range interaction together with the dipolar interaction 
makes the DBEC an attractive system for
experimental soliton generation 
and a challenging system for
theoretical investigation.  
There have
already been studies of bright solitons in DBEC~\cite{dipsol}.

Using the numerical and variational solution of the 
GP equation we predict and study stationary and 
dynamical properties of three-dimensional (3D) DBEC gap solitons, of 
both cigar and disk shapes, localized in the lowest band gap 
by three orthogonal OL potentials. 
The BEC gap 
solitons are realizable only for all repulsive atomic interactions below 
a maximum value so that the chemical potential falls in the band gap. 
The DBEC gap solitons in disk shape can also be 
formed for weakly attractive short-range interaction. The cigar-shaped 
DBEC gap solitons are formed only for short-range repulsion above a 
limiting value. 
We also consider the one-dimensional gap solitons of direct 
experimental interest  
confined by an OL along the dipole
moment direction and harmonic traps in transverse directions. 

We numerically explore the breathing oscillation of the gap solitons upon a 
small perturbation.   We illustrate the dragging of the gap solitons  by an 
OL moving along the axial $z$ direction \cite{MOVEOL,MOVEOL2}. These solitons can be dragged  
without deformation for a reasonably large velocity.

Here we  study the DBEC gap soliton  
of $N$ atoms, each of mass $m$, using the dimensionless 
GP
equation  \cite{pfau}
\begin{eqnarray}  \label{gp3d} 
i  \frac{\partial \phi({\bf r},t)}{\partial t}
 & = & \biggr[ -\frac{1}{2}\nabla^2 +V({\bf r}) + 4\pi a N|\phi({\bf r},t)|^2\nonumber \\
&+ & N \int U_{dd}({\bf r -r'})|\phi({\bf r'},t)|^2d{\bf r'}
\biggr] \phi({\bf r},t), \end{eqnarray} 
with  dipolar interaction 
$
 U_{dd}({\bf R}) = 3
a_{dd}(1-3\cos^2\theta)$ $/R^3,$ $\quad {\bf R=r-r'}.
$
 Here $V({\bf r})$ is the confining potential,
$\phi({\bf r},t)$  the 
wave function at time $t$ with normalization $\int |\phi({\bf r},t)|^2 d {\bf r}=1$, 
 $a$ the atomic scattering length, $\theta$ 
the angle between $\bf R$ and the  polarization direction  $z$.  
The constant $a_{dd}
=\mu_0\bar \mu^2 m /(12\pi \hbar^2)$ 
is a length characterizing the strength of 
dipolar interaction and its experimental
value for $^{52}$Cr  is $15a_0$ \cite{pfau}, with  $a_0$ the Bohr 
radius, 
 $\bar \mu$ the (magnetic) dipole moment of a single atom, and $\mu_0$ 
the permeability of free space. The OL potential in a specific 
direction, say $\hat z$, is $V(\hat z)=s_{\hat z} E_R \sin^2 (2\pi \hat z/
\lambda)$, 
where $E_R=$ $h^2/(2m\lambda^2)$ is the recoil energy of an atom, $\lambda$ 
is the wave length of the laser and $s_{\hat z}$ is the strength of the OL.
In (\ref{gp3d}) length is measured in units of $\lambda/2\pi$
(taken here as 1 $\mu$m for a far infrared laser), time $t$
in units of $m\lambda^2/2\pi h$, and energy in units of $h^2/m\lambda^2$.  
The dimensionless 3D 
periodic OL trap can now be written as  
\begin{equation}
 V({\bf r}) = -[
V_\rho \{ \cos\left(2 x\right)+ \cos\left(2 y\right)\}
+V_z \cos\left(2 z\right)],
\end{equation}
where the parameters $V_\rho$ and $V_z$
are the
strengths of the OL's in radial and axial directions and can be varied
to achieve the disk- ($V_z> V_\rho$) and cigar-shaped
($V_\rho> V_z$) gap solitons in 
DBEC.

{The gap solitons  considered here 
 predominantly have a
Gaussian shape and for these solitons a Gaussian
variational solution is known  to 
lead to a good description \cite{gapfermi}. 
The 
variational approximation provides an analytical understanding  
  and also yields interesting results when the numerical 
procedure is difficult to implement. 
A better but more complicated
description can be obtained with a different ansatz of variational 
functions \cite{auf}.}
 The Lagrangian density of 
  (\ref{gp3d})  is given by
\begin{eqnarray}
{\mathcal L}& = & \,\frac{i}{2}\left( \phi \phi^{\star}_t
- \phi^{\star}\phi_t \right) +\frac{1}{2}\vert\nabla\phi\vert^2
+ V({\bf r})|\phi|^2+ 2\pi aN\vert\phi\vert^4
\nonumber \\ & 
+ & \frac{N}{2}\vert
\phi\vert^2\int U_{dd}({\mathbf r}-
{\mathbf r'})\vert\phi({\mathbf r'})\vert^2 d{\mathbf r}'
.\label{eqn:vari}
\end{eqnarray}
We use the Gaussian ansatz \cite{you} 
\begin{equation}
\phi({\bf r},t)= \frac{\pi^{-3/4}}{w_\rho \sqrt {w_z}} 
\exp\left(-\frac{\rho^2}{2w_\rho^2} - \frac{z^2}{2w_z^2} +i\alpha\rho^2
+i\beta z^2 \right) 
\end{equation}
for a variational calculation, 
where $w_\rho$ and $w_z$ are time-dependent radial and axial widths,
and $\alpha$ and $\beta$  time-dependent phases. 
The effective Lagrangian $L\equiv 
 \int {\mathcal L}d{\mathbf r}$ (per particle) becomes
\begin{eqnarray}
L = &\,
  (w_\rho^2\dot{\alpha} +
w_z^2\dot{\beta}/2) -2V_\rho \exp(-w_\rho^2) -V_z\exp(-w_z^2)
\nonumber \\ 
&\,+
{1}/(2{w_\rho^2}) + {1}/({4w_z^2})
+ 2w_\rho^2 \alpha^2 + w_z^2\beta^2\nonumber \\
&\,+ {N a_{dd}}/{(\sqrt{2 \pi}
w_\rho^2w_z)} [ {a}/{a_{dd}} -
f(\kappa)], \label{lag:eff}
\end{eqnarray}
with 
\begin{eqnarray} && f(\kappa)= [1+2\kappa^2-3\kappa^2d(\kappa)]/(1-\kappa^2),  \\
&& d(\kappa)= \frac{\mbox{atanh}\sqrt{1-\kappa^2}}{\sqrt{1-\kappa^2}}, \;\;
\kappa=\frac{w_\rho}{w_z}.
\end{eqnarray}
The Euler-Lagrange equations for parameters $w_\rho, w_z, \alpha$ and $\beta$ yield
the following equations for widths
 $w_\rho$ and $w_z$ 
 \begin{eqnarray} &
\ddot{w}_{\rho}+\frac{4V_\rho  {w_\rho}}{e^{w_\rho^2}}
=
\frac{1}{w_\rho^3} +\frac{
N}{\sqrt{2\pi}} \frac{  \left[2{a} - a_{dd}
{g(\kappa) }\right]  }{w_\rho^3w_{z}}
,
\label{f1} \\ & \ddot{w}_{z} + \frac{ 4V_z  
{w_z}}{e^{w_z^2}} =
\frac{1}{w_z^3}+ \frac{ 2N}{\sqrt{2\pi}}
\frac{ \left[{a}-a_{dd}
h(\kappa)\right]  }{w_\rho^2w_z^2} , \label{f2} 
\end{eqnarray}
with 
\begin{eqnarray}
&& g(\kappa)=[2-7\kappa^2-4\kappa^4+9\kappa^4 d(\kappa)]/(1-\kappa^2)^2, \\ 
&& h(\kappa) =[1+10\kappa^2 -2\kappa^4 -9\kappa^2 d(\kappa)]/(1-\kappa^2)^2.
\end{eqnarray}
The chemical potential $\mu$ is given by 
\begin{eqnarray}
\mu& = & \frac{1}{2w_\rho^2}+
\frac{1}{4w_z^2}+\frac{2N[a-a_{dd}f(\kappa)]}{\sqrt{2\pi} w_zw_\rho^2} \nonumber \\
 & & -2V_\rho \exp(-w_\rho^2) -V_z\exp(-w_z^2).
\end{eqnarray}

For gap solitons the system must be repulsive. For a normal 
BEC ($a_{dd}=0$), attraction corresponds to $a<0$ and repulsion to 
$a>0$ and gap solitons are formed below a limiting repulsive scattering 
length $0<a<a_c$.  For $a_{dd}>0$, in the disk shape, the dipole 
moment contributes repulsively   and {due to this extra repulsion} 
a DBEC gap soliton can be formed in a 
window of scattering lengths $-a_1<a<a_2$ between a limiting attractive 
($-a_1$) and repulsive ($a_2$) limits.  
However, in a weak cigar shape, 
the dipole moment contributes attractively  and {due to the extra attraction}
a DBEC gap soliton can be 
formed in a window of repulsive scattering lengths  $a_3<a<a_4$. 
{
The limiting values $a_1, a_2, a_3$ and $a_4$ can be obtained from a solution of the GP 
equation.   }
In a  strong cigar regime, the dipole moment contributes to strong 
attraction and no gap soliton can be formed due to collapse 
instability.

The cigar-shaped quasi-1D gap solitons with a 
radial  harmonic trap of frequency $\Omega_\rho$ 
satisfy  (\ref{gp3d}) with $V({\bf r})
= \Omega_\rho^2 \rho^2/2-$ $ V_z\cos(2z)$. In this case it is convenient 
to solve the 1D GP equation \cite{1D,1D2} with the reduced dipolar interaction 
\begin{eqnarray}\label{1dpot}
U_{dd}^{1D}(Z)=&\, 3a_{dd}N[4\delta(\sqrt t)/3+ 2 \sqrt t-\sqrt \pi
(1+2t)\nonumber \\
&\, \times e^t \{1- \mbox{Erf}(\sqrt t)  \}]/(\sqrt 2d_\rho )^3, 
\end{eqnarray} 
where $Z=\vert z-z'\vert$,  
$d_\rho= 1/\sqrt {\Omega_\rho}$, $t=[Z/(\sqrt 2 d_\rho)]^2$ and where we have 
included the proper $\delta$-function  term mentioned in \cite{1D}.
The 1D GP equation is given by
\begin{eqnarray}  \label{gp1d} 
i  \frac{\partial \phi_{1D}(z,t)}{\partial t}
 & = & \biggr[ - \frac{1}{2}\frac{\partial^2}{\partial z^2} -V_z\cos(2z) + 
\frac{2 a N}{d_\rho^2}|\phi_{1D}(z,t)|^2\nonumber \\
& & + N\int U_{dd}^{1D}( Z )\vert\phi_{1D}({z'},t)\vert^2d{z'}
\biggr] \phi_{1D}(z,t), 
\end{eqnarray} 
Using the Gaussian ansatz 
\begin{equation}
\phi_{1D}(z)=\frac{\pi^{-1/4}}
{\sqrt{w_z}} \exp\left[-\frac{z^2}{2w_z^2}+i\beta z^2\right],
\end{equation} 
the variational Lagrangian 
becomes 
\begin{eqnarray}
L & = & \frac{w_z^2\dot{\beta}}{2} 
 -V_z\exp(-w_z^2)+ \frac{1}{4w_z^2}
+ w_z^2\beta^2 \nonumber \\
& & + \frac{N a_{dd}}{\sqrt{2 \pi} d_\rho^2w_z} \left[ \frac{a}{a_{dd}} -
f(\kappa_0)\right]; \quad  \kappa_0=\frac{d_\rho}{w_z}.
\end{eqnarray}
The variational 
equation for $w_z$ becomes
\begin{eqnarray}\ddot w_z + \frac{4V_zw_z}{\exp(w_z^2)} = \frac{1}{w_z^{3}}
+\frac{2 N[a-a_{dd}h(\kappa_0)]}
{\sqrt
{2\pi}w_z^2 d_\rho^2}. \end{eqnarray}
The corresponding 
chemical potential is 
\begin{eqnarray}\mu=
\frac{1}{4w_z^2}+\frac{2N[a-a_{dd}f(\kappa_0)]}{\sqrt{2\pi} w_z d_\rho^2} 
 -\frac{V_z}{\exp(w_z^2)}. \end{eqnarray}

\begin{figure}
\begin{center}
\includegraphics[width=\linewidth]{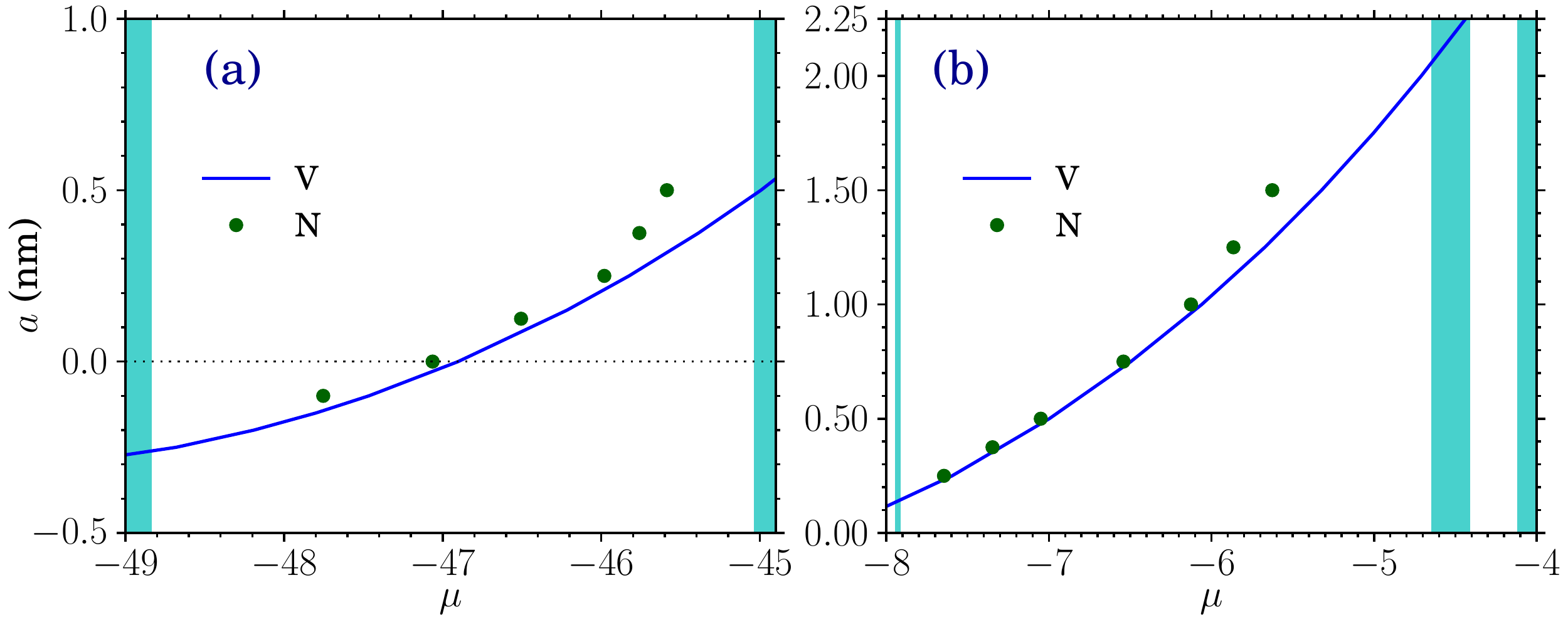}
\end{center}

\caption{(Color online) Band (shaded area),  gap (white area)  
 and the  numerical (N) and variational (V)
 $a-\mu$ plots  of 3D DBEC gap
solitons for $a_{dd}=15a_0$,  $N=500$
and $V_\rho = 5,$ (a) $V_z$ = 50 (disk) and (b) 
$V_z =4$ (cigar).}

\label{fig1}
\end{figure}

\begin{figure}[!b]
\begin{center}
\includegraphics[width=\linewidth]{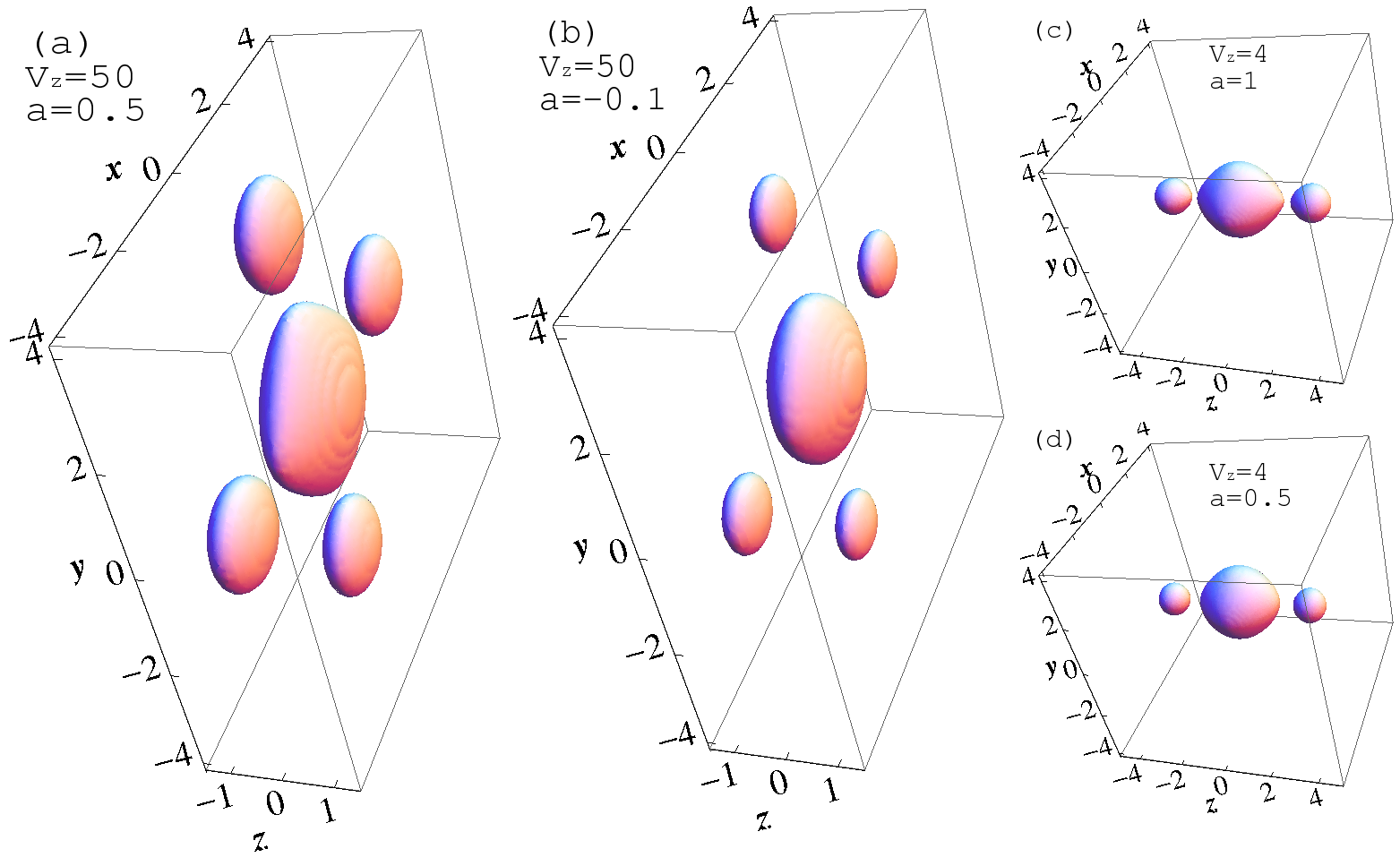}
\end{center}

\caption{(Color online) 3D contour density plot of   
gap solitons for $N=500, V_\rho=5, a_{dd}=15a_0$. (a)
$V_z=50, a=0.5$, (disk), (b) $V_z=50, a=-0.1$, (disk), 
(c) $V_z=4, a=1$, (cigar), (d) $V_z=4, a=0.5$, (cigar). 
The density at  contour is 0.002.
}

\label{fig2}
\end{figure}

To study the gap solitons, 
we perform a 3D numerical simulation
employing real-time propagation with Crank-Nicolson 
method \cite{CPC}. {The numerical solutions are 
obtained  
by  averaging the wave function
over iterations with the variational solution as input.}
The dipolar interaction is
evaluated by  fast Fourier transform \cite{jb}.

The stationary gap solitons appear in the band gap of the OL and a 
prior knowledge 
of the band and gap of the periodic OL is of advantage. The dipolar 
interaction is most prominent in the cigar (attractive) and disk 
(repulsive) shapes and hence we study the gap solitons in these two 
shapes. The OL parameters in these cases are taken as (a) $V_\rho = 5, 
V_z$ = 50 (disk shape) and (b) $V_\rho = 5, V_z$ = 4 (cigar shape).  The 
band and gap in these cases are shown in figure  \ref{fig1} together with 
the variational and numerical chemical potential $\mu$ for the stable gap 
solitons in the lowest band gap as a function of scattering length $a$ for a DBEC 
($a_{dd}=15a_0$). The gap solitons in the higher band gaps are found to be 
dynamically unstable. 
In figure  \ref{fig1},
 the agreement between variational and numerical
 results
worsens 
for large $a$. The system is more repulsive for large $a$ and tends to occupy 
multiple OL sites and a single-peak Gaussian variational ansatz may not 
provide a good approximation to reality. In the cigar shape ($V_z=4$), 
the dipolar interaction is attractive and hence DBEC gap solitons are 
{possible for $a\equiv a_3 >0.1343 $ nm as in figure  \ref{fig1} (b). In the disk shape 
($V_z=50$), the dipolar interaction is repulsive, and hence DBEC gap 
solitons are possible for $a\equiv -a_1  > -0.2635 $ nm as in figure  \ref{fig1} (a).  }

\begin{figure}
\begin{center}
\includegraphics[width=\linewidth]{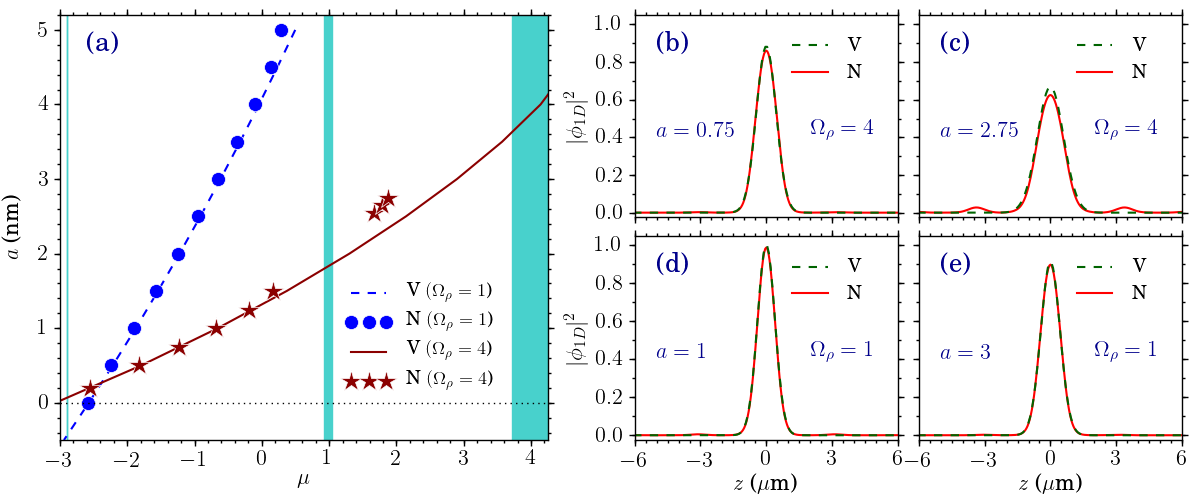}
\end{center}

\caption{(Color online) (a)  Band (shaded area),  gap (white area)  
 and the numerical (N) and variational (V) $a-\mu$ plot of 1D DBEC gap 
solitons for $\Omega_\rho$ = 1 (disk) and $\Omega_\rho = 4$ (cigar).  
Variational and numerical densities of the 1D DBEC for (b) 
$\Omega_\rho=4, a=0.75$ nm, (c) $\Omega_\rho=4, a=2.75$ nm, (d) 
$\Omega_\rho=1, a= 1$ nm, and (e) $\Omega_\rho=1, a=3$ nm. In all cases 
$a_{dd}=15a_0$, $N=500$ and $V_z = 5.$
}

\label{fig3}
\end{figure}

In figure  \ref{fig2} we show the 3D contour density profiles of some typical
stationary gap solitons with $N=500, V_\rho=5, a_{dd}=15a_0$ for variable 
$a$ and $V_z$. We show disk-shaped profiles for  $V_z=50$ and (a)
$a=0.5$ and (b) $a=-0.1$ and cigar-shaped profiles for $V_z=4$ and (c) $a=1$ 
and (d) $a=0.5$. For both disk and cigar shapes, the increased scattering length 
implies more nonlinear repulsion and  one can have more atoms in adjacent OL 
sites. {Although the gap solitons in figure  \ref{fig2} occupy multiple OL 
sites, more than 95 $\%$ of the matter is contained in the central site and an approximate 
Gaussian distribution is valid.}

\begin{figure}[!b]
\begin{center}
\includegraphics[width=\linewidth]{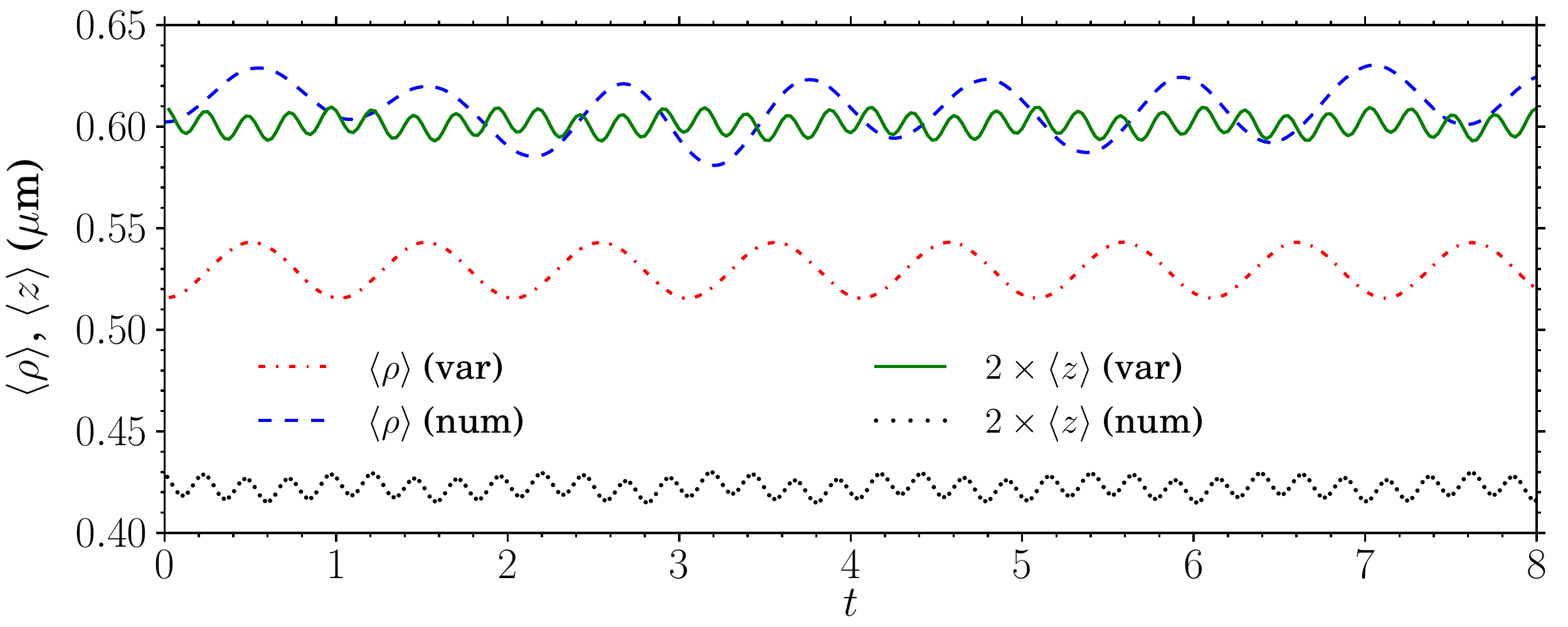}
\end{center}

\caption{(Color online) Numerical (num) and variational (var) 
rms sizes $\langle \rho \rangle, \langle z \rangle$ 
during breathing oscillation of a disk-shaped 3D DBEC for $N=500, a_{dd}=
15a_0, V_z=50, V_\rho =5$ as $a$ is changed from 0 to 0.1.
}

\label{fig4}
\end{figure}

Next we study 
1D gap solitons trapped by the OL $-V_z \cos(2z)$ along the $z$ direction 
and harmonic trap $\Omega_\rho^2 \rho^2$ 
in transverse directions as calculated 
from  numerical and  variational solutions of the reduced 
1D GP equation \cite{1D} with dipolar interaction (\ref{1dpot}). 
 In figure  \ref{fig3} (a) we show the bands
 and gaps as well as the $a-\mu$ plot of the 1D gap solitons. For 
$\Omega_\rho =1 ,$ the DBEC is of disk shape and the contribution of the 
dipolar term is repulsive, and  for $\Omega_\rho =4 ,$ DBEC is of cigar shape 
and the dipolar term contributes attractively.  
 In figures 
\ref{fig3} (b) $-$ (e) we plot the typical numerical 1D density profiles 
of some gap solitons and compare with variational results.  Although, variational 
results exist across the bands, stable DBEC 3D gap solitons exist in the 
lowest band gap  away from the band. In 1D some stable gap solitons are found in the 
first excited gap.
Gap solitons cannot be stabilized  
close to the bands, e. g., near $\mu=-45$ and $-49$ in figure  
\ref{fig1} (a), near $\mu=-4.5$ and $-8$ in figure  \ref{fig1} (b), and near $\mu=1$ 
and 4 in figure  \ref{fig3}. 
In  figures \ref{fig1} (a) and \ref{fig3} (a)
we clearly see that gap solitons are possible for small negative scattering lengths in 
disk-shaped DBEC and that, in the second band gap,  1D gap solitons can be stabilized 
  only in a small domain of $\mu$ values between two bands.

To test the stability of the 3D DBEC gap solitons we now consider the 
breathing oscillation of a disk-shaped soliton initiated by slightly changing the 
scattering length $a$. Such a change in $a$ can be made by varying 
a background magnetic field near a Feshbach resonance \cite{fesh}. In figure  
\ref{fig4} we plot the variational and numerical  rms sizes $\langle \rho \rangle, \langle z 
\rangle$ versus time $t$.   We find, using Fourier
analysis,  that the principal axial 
frequencies  1.0137 (variational) and 1.0812 (numerical) and the 
principal radial frequencies  0.2421 (variational) and 0.2469 
(numerical) are in good agreement with each other.

\begin{figure}[!t]
\begin{center}
\includegraphics[width=\linewidth]{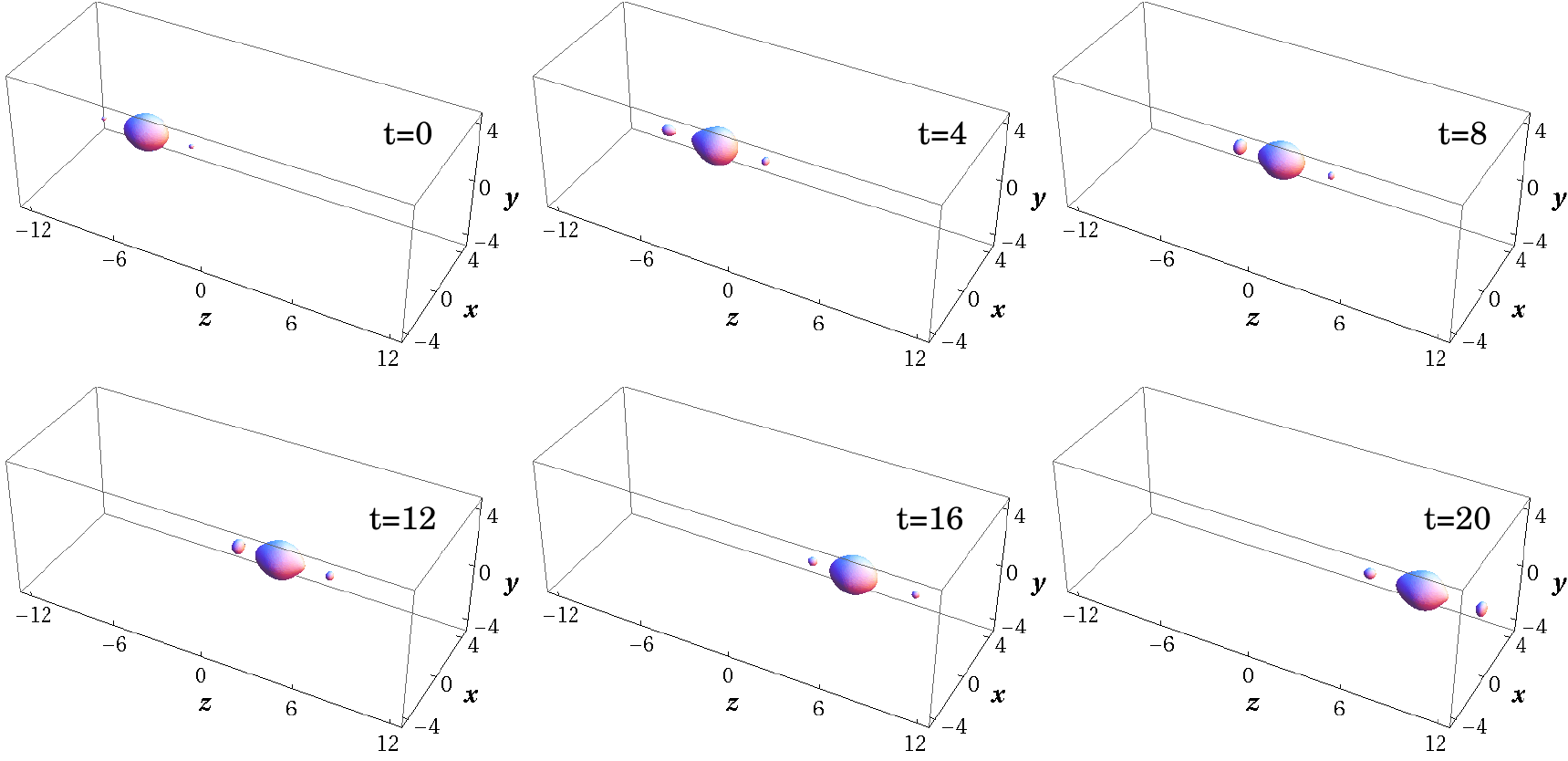}
\end{center}

\caption{(Color online) 
Snapshots of 3D contour density profiles of a cigar-shaped gap soliton 
for $V_z=4, V_\rho =5, N=500, a_{dd}=15a_0, $ and $a= 0.5$ nm during 
dragging by the moving OL $-V_\rho[\cos(2x)+\cos(2y)]-2V_z\cos[2(z-vt)], 
v=0.75$ 
at times $t=0,4,8,12,16
$ and 20. 
}

\label{fig5}
\end{figure}

Finally, we  study the stability of a cigar-shaped 3D DBEC when dragged 
by an OL moving in the axial $z$ direction. It has been experimentally 
found that a BEC confined by a transverse harmonic trap 
remains stable \cite{MOVEOL2} when dragged by a moving 
1D OL along the axial $z$ direction below a critical velocity $v_c$ of 
half the recoil velocity [$v_R=h/(m\lambda)$],
which in present  units ($\lambda=2\pi$ and $m=\hbar=1$) is $v_c=v_R/2=0.5$. 
Similar result  is 
found to be true in the case of present 3D DBEC
gap solitons.  Steady dragging is possible 
in this case even for velocities slightly larger than this limit. 
Instant snapshots of dragging dynamics with a 
velocity $v=0.75 (<v_c=1)$ along $z$ direction 
is 
shown in figure  \ref{fig5} where we find that the gap soliton can be dragged without 
distortion with a reasonably large velocity for a considerable 
time. A movie of dragging of soliton of figure \ref{fig5} showing stability is prepared and contained in supplementary clip S1 
(see supplementary data available at stacks.iop.org/JPhysB/44/xxxxxx/mmedia, also available at 
 http://www.youtube.com/watch?v=wJbbKmsjQuU).

To conclude, we suggested the possibility of 3D DBEC gap solitons
of about 1000  Cr atoms 
confined in the lowest band gap 
by three OL  in orthogonal directions and studied their 
statics (shape and chemical potential) and dynamics (breathing 
oscillation and dragging by an OL). In addition we studied 1D DBEC gap 
solitons using reduced 1D GP equations with a transverse harmonic trap 
and an axial OL  along polarization direction. The 3D DBEC gap solitons are 
found to be dynamically stable during breathing oscillation and dragging 
for a long enough time  for experiments and,  with available 
technology, these solitons could be created and studied  in laboratory.
The present study opens 
up new directions of research that include, among others,
excited states of gap solitons \cite{10r} 
and vortex gap solitons \cite{vgap}
 in DBEC.


\ack
We thank FAPESP (Brazil), CNPq (Brazil), DST (India),
and CSIR (India) for partial support.

\end{document}